\documentclass[epj, nopacs]{svjour}

\usepackage[numbers,sort&compress]{natbib}

\usepackage{amsmath}
\usepackage{graphicx}
\usepackage{subfigure}
\usepackage{comment}
\usepackage{soul}
\usepackage{float}
\usepackage{slashed}
\usepackage{pbox}
\usepackage{multirow}
\usepackage{color}  

\newcommand{\be}{\begin{eqnarray}}
	\newcommand{\ee}{\end{eqnarray}}

\def\beq{\begin{equation}}
	\def\eeq{\end{equation}}
\def\bit{\begin{itemize}}
	\def\eit{\end{itemize}}
\newcommand{\n}{\nonumber}

\definecolor{orange}{rgb}{1,0.5,0}

\begin{document}

	\title{Relativistic Particle on Light-Front}
	\author{Junmou Chen
		\thanks{\emph{chenjm@jnu.edu.cn}}
	}     
	\institute{Department of Physics, School of Physics and Engineering, Jinan University, Guangzhou,  China}
	\date{Received: date / Revised version: date}
	
	\abstract{	We construct one-particle states as unitary, irreducible representations of Poincare group in front form, characterized by a special null vector, dubbed the reference vector.  We demonstrate that this construction has massive-massless continuation. 	 The state is defined by the reference vector.  The little group transformation, defined at a general moving momentum,  is proved to be  equivalent to a change of reference vector.  The resulting Wigner D-matrix is parameterized by the rapidities, in addition to the two reference vectors before and after transformation.  Boosting the rapidities to infinity,  it reaches the  massless limit smoothly. We then apply those results to massive spin-1 particle and compute the corresponding  Wigner D-matrix.  The resulting polarization vectors are equivalent to those in spinor-helicity formalism. In the massless limit,  it is shown that longitudinal polarization decouples from the spectrum.  The  $\epsilon^\mu_\pm \rightarrow \epsilon^\mu_\pm +\xi k^\mu$ shift turns out to be remnant of this decoupling, with $\xi$ determined by the angle between the reference vectors.  Our results thus give us a new perspective on gauge symmetry:   it can be understood as  the equivalence between different forms of massless spin-1 polarization vectors from reaching the massless limit through  different reference vectors. 
	} 

\makeatletter
\def\makeheadbox{}
\def\@date{}  
\makeatother

	\maketitle		


	\section{Introduction}
	\label{sec:intr}

	Ever since Wigner's work of classifying relativistic quantum particle states as unitary, irreducible representations of Poincare group was published \cite{Wigner:1939cj,Bargmann:1948ck, Weinberg_1995}, it has been a cornerstone for  quantum field theory.  Traditionally  massive particle and massless particle are treated separately according to  their respective little groups. In particular for the massive particle, the little group transformation and Wigner's D-matrix elements are  computed in the rest frame.  This construction, however, has a deficiency that one cannot obtain little group transformation of massless particle from that of  massive particle by taking the mass to $0$, as one would naively expect. 
	
	This is unsatisfying in light of the fact that one can directly prove this massive-massless continuation in Lie algebra using  group contraction. Making a change of variable to the Lie algebra of the little group $SO(3)- \{J_1,J_2,J_3\}$  for massive particle: $J_1\rightarrow S_1=\epsilon J_2, J_2\rightarrow S_2= \epsilon J_1$, the Lie algebra becomes
	\be\label{eq:grp_contr}
	[S_1, S_2]=i \epsilon^2J_3, \  \ [S_2,J_3]=i  S_1, \ \ [J_3,S_1]=i  S_2
	\ee
	When $\epsilon\rightarrow 0$, the above equations become the Lie algebra of $ISO(2)$, precisely the little group for massless particle.   Despite the proof at the level of Lie algebra, however,  there hasn't been a full construction of unitary representation of Poincare group that satisfies massive-massless continuation as illustrated in Eq.(\ref{eq:grp_contr}).  The goal of this paper is to provide one.  
	
	Our motivations are not merely formal.  In recent decades, there has been enormous advance in the research of scattering amplitudes for  massless  particles \cite{Dixon:2013uaa, Elvang:2013cua, Parke:1986gb, Berends:1987me, Dixon:1996wi, Mangano:1990by,  Bern:1996je,  Britto:2004ap, Britto:2005fq, Bern:1991aq, Bern:1994cg, Bern:1994zx, Bern:2010ue, Bern:2005iz, Bern:2008qj, Bern:2019prr, Witten:2003nn, Arkani-Hamed:2008owk, Arkani-Hamed:2012zlh, Arkani-Hamed:2013jha, Cachazo:2014xea, Drummond:2009fd}. In comparison, even though there has also been much progress for massive amplitudes too \cite{Arkani-Hamed:2017jhn, Conde_2016, Gomez-Laberge:2025yid, Wu_2022, Ni:2025xkg, Ema:2024vww, Ballav_2021}, they are far less developed than the massless counterparts.  One of the main sources of the advance in massless amplitudes is the use of little group transformation in constraining amplitudes.  Thus, it seems reasonable to hope a deeper understanding of little group transformation on massive particles would help us advance massive amplitudes as well.  Especially, a construction of Wigner's classification with massive-massless continuation would enable us to make use of all the established results in the massless case, thus has the potential of enormous benefits.  We will particularly focus on spin-1 particle, which has a gauge symmetry in the massless limit from the little group.  We also hope that a construction of massive-massless continuation can shed new lights on our understanding  of gauge symmetry.

	Another related motivation  is to derive the general Lorentz transformations of  wave functions  in spinor-helicity formalism, which we notice are not the standard ones in textbooks.  Wave functions  in spinor-helicity formalism have a particular spatial direction defined by $n$ (see Appendix(\ref{sec:append})). This can be seen most clearly in massless spin-1 particles, for which  the polarization vectors   can be obtained by imposing the conditions of $k\cdot \epsilon=0$ and $n\cdot \epsilon =0$ with $n^2=0$ \cite{Dixon:1996wi}.  Those wave functions cannot be derived by Lorentz symmetry as commonly defined, which classifies rotations and boosts separately,  as it seems to break rotational symmetry.  Nevertheless, as classified by  Dirac \cite{Dirac:1949cp, Brodsky:1997de},  there are three different forms of Lorentz group.  While one of them is the  form  commonly used in the literature --  dubbed instant form, there  is another form called front form, which can be understood to classify the Lorentz generators according to a special null direction.  It's natural to conjecture that the front form of Lorentz group can be the basis for wave functions defined by a null vector $n$. Indeed, the wave functions for spin-$\frac{1}{2}$ and spin-1 in  front form have been derived  in \cite{Chiu:2017ycx}. What's missing is how they transform under general Lorentz transformations. 
	
	Front form also gives a natural way to implement massive-massless continuation. Instead of reducing Lorentz transformations of a particle state to  the rest frame as in instant form, in front form  Lorentz transformations of a particle can be reduced to  the null reference vector(as will be shown in the paper), without fixing to a particular reference frame. This gives the opportunity to reach the massless limit through  boosting to infinite rapidity.   
	In conclusion, based on the analysis above, we will  construct particles as unitary representations of Poincare group based on the front form,  which turns out to have smooth massless limits.  The remainder of the paper is organized as follows:
	
	In Sec.(\ref{sec:frt_form}), we give a very brief review of Poincare group in front form, focusing on the elements related to our construction. 
	
	In Sec.(\ref{sec:construction}),  we construct the one particle state on a moving frame in front form,  labeled by a reference null vector $n$, as well as the rapidity $\eta$ along this direction.  The little group transformation turns out to be  equivalent to a simple change of reference vector. As a result, the Wigner D-matrix elements equal to the inner product of states with different reference vectors.  We then derive a concrete formula to calculate those Wigner D-matrix/d-matrix elements. 
	
	In Sec.(\ref{sec:spin-1}), we apply our construction and formulae to spin-1 particle, and calculate the corresponding Wigner D-matrix elements. The results agree with the conventional Wigner D-matrix elements in the rest frame.   Taking the massless limit, the little group transformation of spin-1 particle reduces to on-shell gauge symmetry: $\epsilon\rightarrow \epsilon+\xi k$, with the coefficient $\xi$  determined by the angle between the two different reference vectors. 
	
	We also compute the Wigner D-matrix elements from the Lie algebra of the little group directly.  The results agree with our construction in front form when the parameters are matched with each other,  providing   cross check and confirmation for our results.

	Finally, in Sec(\ref{sec:con}) we conclude and discuss some physical implications of our results. 
	
	We also have an appendix to explain in details the connection between polarization vectors of spin-1 particle in front form and in spinor-helicity formalism.

	\section{Poincare Group in Front Form}
	\label{sec:frt_form}
	
	
	The Poincare group in front form is defined by setting the time variable as $x^{\pm}=x^0\pm x^3$ and $x_{\pm}=\frac{1}{2}(x^0\pm x^3)$ \cite{Brodsky:1997de}, which can be also written as $x^+=x\cdot n$ with null vector $n^\mu=(1,0,0,-1)$.   It's most convenient to use light-cone coordinates in front form,  with $x^\mu=(\underset{+}{x^+},\underset{-}{x^-},\underset{1}{x^1},\underset{2}{x^2})$, $x_\mu=(\underset{+}{x_+},\underset{-}{x_-},\underset{1}{x_1},\underset{2}{x_2})=(\underset{+}{\frac{x^-}{2}},\underset{-}{\frac{x^+}{2}},\underset{1}{-x^1},\underset{2}{-x^2})$.  Here the labels $+-12$ under the components denote  light-cone coordinates, as will be the case for the remainder of the paper.  On the other hand, when there is no label under the components, it means the ordinary coordinates in instant form.

	The  generators of Poincare group in front form are also classified according to the null direction $n$:
	\be
	P^\mu&=&(P^+,P^-,P^1,P^2)\\
	M^{\mu\nu}&=&\{M^{+-},M^{12}, M^{\pm 1}, M^{\pm2} \} 
	\ee
	with $P^{\pm}\equiv P^0\pm P^3$, $M^{+-}\equiv M^{03}$, $M^{\pm i}\equiv M^{0i}\pm M^{3i}$ with $i=1,2$. 
	
	All nonzero commutation relations of the Lie algebra are listed as below:
	\begin{eqnarray}
		[M^{+-}, P^\pm] &=& \mp i P^\pm, \ \   [M^{+-},M^{\pm i}] = \mp iM^{\pm i}   \nonumber \\
		\left[ M^{12}, M^{\pm i}\right]&=&i\epsilon^{ij}M^{\pm j}, \ \ [M^{12}, P^i] =i\epsilon^{ij}P^j   \\
		\left[P^{\pm}, M^{\mp i}\right]&=& 2iP^i, \ \ [P^i, M^{\pm j}]=iP^\pm\delta^{ij},  \nonumber \\
		\left[M^{+i}, M^{-j}\right]&=&-2iM^{+-}\delta^{ij}-2i\epsilon^{ij}M^{12} \nonumber 
	\end{eqnarray}
	From the commutation relations, we can compute and obtain the corresponding  Lorentz transformations.  
	$M^{+-}$ generates the longitudinal boost,  under which the components of $P^\mu$ undergo simple rescalings \cite{Kogut:1969xa}:
	\be\label{eq:long_boost}
	e^{i M^{+-}\eta}P^\pm e^{-iM^{+-}\eta} &=& P^\pm e^{\pm \eta}  \\
	\ \  \ e^{i M^{+-}\eta}P^i e^{-iM^{+-}\eta} &=& P^i   \nonumber 
	\ee

	$M^{+ i}$ generates transverse boosts, under which the components of $P^\mu$ transform as  
	\be\label{eq:tran_boost}
	e^{+i M^{+i}\theta^{+i}}P^\pm e^{-iM^{+i}\theta^{+i}} &=& P^\pm +\delta^{\pm-}( 2\theta^{+i}P^i + (\theta^{+i})^2P^+)
	\nonumber \\
	e^{i M^{+i}\theta^{+i}}P^j e^{-iM^{+i}\theta^{+i}} &=& P^j + \theta^{+i}P^+
	\ee
	with no summation for repeated $i$ indices.  The vector representations of the generators can be derived straightforwardly from Eq.(\ref{eq:long_boost}), Eq.(\ref{eq:tran_boost}) and momentum $p^\mu$ as the eigenstate of $P^\mu$. 
	

	Based on the information above, Lorentz transformation from momentum  $k^\mu=(\underset{+}{k^+},\underset{-}{k^-},\underset{1}{k^1},\underset{2}{k^2})$ to  momentum $p^\mu=(\underset{+}{p^+},\underset{-}{p^-},\underset{1}{p^1},\underset{2}{p^2})$ can be achieved by a longitudinal boost, followed by a transverse boost and then a longitudinal rotation 
	\be\label{eq:stl}
	k^\mu \rightarrow p^\mu =  e^{iM^{12}\varphi}e^{+iM^{+1}\theta^{+1}}e^{iM^{+-}\eta} \  k^\mu
	\ee
	In the special case of $k^\mu=(\underset{+}{m},\underset{-}{m},\underset{1}{0},\underset{2}{0})$, the parameters can be expressed by  $p^\mu$ alone 
	with $e^{-\eta}=\frac{m}{p^+}, \theta^{+1}= \frac{p^T}{p^+}, \cos\varphi = \frac{p^1}{p^T} $, $p^T=\sqrt{(p^1)^2+(p^2)^2}$. 
	
	We call this transformation  the standard Lorentz transformation, denoting  as $L(p,k,n)$:
	\begin{equation}\label{eq:stl_1}
		L(p,k,n) = e^{+iM^{12}\varphi}e^{+iM^{+1}\theta^{+1}}e^{+iM^{+-}\eta}
	\end{equation}
	Notice we have  labeled $L(p,k,n)$ with the null vector $n$, which specifies the direction of the longitudinal boost and defines the front form.   However, since the direction of $n$ is purely conventional, we should be able to  define the standard Lorentz transformation with any null vector.  This freedom in choosing  the null vector that defines front form  will be crucial for constructing unitary representations of Poincare group in Sec.(\ref{sec:construction}). 
	There is, however, a problem, which is  it's not clear how to calculate Lorentz transformations defined on a null vector that's not $n^\mu= (1, 0, 0, -1)$.     We will give a recipe on how to do it  in Sec.(\ref{subsec:wig_d}).

	\section{One Particle State in Front Form and Wigner D-matrix}
	\label{sec:construction}

	\subsection{General Lorentz Transformation of One Particle State } 
	\label{subsec:1_ptcl_stt}

	According to Wigner's theorem,   one particle quantum states  as unitary representations of Poincare group are classified by the eigenvalues of the two Casimir operators $P^2=P^\mu P_\mu$ and $S^2=S^\mu S_\mu$, with $P^\mu$ being the momentum generator and $S^\mu=\epsilon^{\mu\nu\rho\sigma}M_{\nu\rho}P_\sigma$ called Pauli-Lubanski pseudo-vector. Their eigenvalues are mass $m$ and spin $s$ respectively.  In addition, we also need  the eigenvalue of $P^\mu$(momentum $p$) and  the eigenvalue of $S^3$ ($\sigma$) to fully specify the state.  A general one particle state can then be written as $|m,p; s,\sigma\rangle$.  For convenience, we usually neglect $m$ and $s$,  writing it as  $|p,\sigma\rangle\equiv |m,p; s,\sigma\rangle$.  In the front form,  we need an additional label.  Generally  $p$ is a  moving momentum,  $|p,\sigma\rangle$ is related to the state in rest frame $|k,\sigma \rangle$  by a standard Lorentz transformation $U(L(p,k,n))$ (Eq.(\ref{eq:stl})), which is labeled by a null vector $n$.  The state $|p, \sigma \rangle$ should then also be labeled by $n$: 
	\be\label{eq:stl_state}
	|p,\sigma, n\rangle \equiv U(L(p,k,n))|k, \sigma\rangle
	\ee
	We will call $n$ the reference vector for the state from now on, which is  the crucial difference between the front form and the instant form. 
	
	Now we will derive the general Lorentz transformation on $|p,\sigma, n\rangle$ by reducing it to the little group. However, in order to facilitate massive-massless continuation, the little group transformation is defined on  a moving frame, instead of the rest frame in the traditional approach.  Therefore we first define another moving momentum $q$ and  the corresponding state $|q,\sigma, n\rangle$ in accordance to Eq.(\ref{eq:stl_state}): 
	\begin{equation}
		|p,\sigma, n\rangle =   U(L(p, q, n)) |q, \sigma, n\rangle 
	\end{equation}
	
	Now acting on $|p,\sigma, n\rangle$ with  a general Lorentz transformation $U(\Lambda)$, we have 
	\begin{eqnarray}\label{eq:Lambda_p}
		&&U(\Lambda) |p,\sigma, n \rangle   \\
		&=&U(L(\Lambda p, q, n))  U(L^{-1}( \Lambda p,q, n) \Lambda L(p,q,n))|q,\sigma,n\rangle  \n 
	\end{eqnarray}
	$W(q; p, \Lambda)\equiv L^{-1}(\Lambda p,q, n)\Lambda L(p,q,n)$ is the little group acting on $q$, under which $ q$ remains unchanged:  $W q=q$.   Acting the little group transformation on  $|q,\sigma ,n \rangle$ we have 
	\be
	U(W(q; p,\Lambda))  | q,\sigma,n \rangle = \sum_{\sigma'}D_{\sigma\sigma'}(q; p, \Lambda)|q,\sigma', n\rangle
	\ee\label{eq:lg_tran_0}
	$D_{\sigma\sigma'}$ is Wigner D-matrix, which is defined as  
	\be
	D_{\sigma\sigma'}(q;p, \Lambda) = \langle q,\sigma', n| W(q; p, \Lambda)|q,\sigma, n\rangle
	\ee
	Plugging in Eq.(\ref{eq:Lambda_p}), we get 
	\begin{eqnarray}\label{eq:gener_tran_1}
		U(\Lambda) |p,\sigma, n \rangle  &=& U(L(\Lambda p, q, n) U(W(q; p, \Lambda))) |q,\sigma,n\rangle  \nonumber \\
		&=& \sum_{\sigma'}D_{\sigma\sigma'}(q; p, \Lambda)U(L(\Lambda p, q, n)) |q,\sigma', n\rangle        \nonumber \\
		\Rightarrow   U(\Lambda) |p,\sigma, n \rangle  &=&\sum_{\sigma'}D_{\sigma\sigma'}(q;p, \Lambda) |\Lambda p,\sigma', n\rangle
	\end{eqnarray}
	Now we have reduced the general Lorentz transformation $\Lambda$ on $|p,\sigma, n \rangle$ to the little group transformation on $q$, which gives the corresponding Wigner D-matrix elements.  

	Then we   give a  precise definition of the little group transformation that's suitable for proper parameterization and calculation.      We define the little group as  
	\be\label{eq:lg_def_1}
	W(q,n|n')=W(q; p, \Lambda) = L(q,k,n')L(k,q,n),
	\ee
	$L(k,q,n) =L^{-1}(q,k,n)$, $k$ is the momentum at the rest frame, $n'$ and $n$ are  two different reference vectors.  First of all,  by fixing $n$ and rotating $n'$,   the direction of $n'$ relative to $n$  covers the whole surface of a 2-sphere($S^2$).  Thus the group defined by Eq.(\ref{eq:lg_def_1}) is precisely the little group of SO(3).  Secondly,   by relating to the rest frame, we can compare our results with the standard construction in instant form conveniently.  Applying Eq.(\ref{eq:lg_def_1}) to $|q,\sigma, n\rangle $, we have
	\begin{eqnarray}\label{eq:lg_tran_1}
		&&U(W(q, n|n') )| q, \sigma, n\rangle  \nonumber \\
		&=& U(L(q,k,n')) U(L^{-1}(q,k,n))| q,\sigma, n\rangle \nonumber \\
		& = & U(L(q,k,n'))| k,\sigma\rangle\nonumber \\
		&\Rightarrow&   U(W(q, n|n') )| q, \sigma, n\rangle  =  |q,\sigma, n'\rangle 
	\end{eqnarray}
	
	So the role of  a little group transformation on one particle state  is to change its reference vector!  Moreover,  the inversion of  $W(q,n|n')$ can also be obtained by  simply exchanging the reference vectors:  $W^{-1}(q,n'|n)= W(q,n|n')$.
	
	Following the discussions above,  $n|n'$ with $n'$ rotating around $n$ could be parameterized by spherical angles $\theta, \varphi$.  However, the spherical angles are not enough to completely fix $W(q,n|n')$. This can be seen  in the definition of  Eq.(\ref{eq:lg_def_1}).   The transverse Lorentz transformations and longitudinal rotations in $L(q,k,n')L(k,q,n)$ are parameterized by $\theta, \varphi$, but the longitudinal boosts are still not accounted for.   They can be parameterized by rapidities $\eta=\ln \frac{n\cdot q}{m}$ and $\eta'=\ln \frac{n'\cdot q}{m}$.  Notice in this parameterization the massless limit of $m\rightarrow 0$ is equivalent to the infinite boost limit of $\eta,\eta'\rightarrow \infty$.  In summary, the little group $W(q, n|n')$ can be parameterized as  
	\[
	W(q,n|n') = W(\eta, \eta', \theta, \varphi) = W(\eta,\eta', n|n')
	\]
	To keep the information of reference vectors explicit,  we simply write the little group  as $W(\eta,\eta', n|n')$ from now on.  Same for the subsequent Wigner D-matrix.  
	
	Combining   Eq.(\ref{eq:lg_def_1}) and Eq.(\ref{eq:lg_tran_1}),  the general Lorentz transformation on one particle state can be finally written as
	\begin{eqnarray}\label{eq:gener_tran_final}
		U(\Lambda ) |p,\sigma, n\rangle    &=&  \sum_{\sigma'
			oo}D_{\sigma\sigma'}(\eta, \eta', n|n') |\Lambda p,\sigma', n\rangle
	\end{eqnarray}
	with  $D_{\sigma\sigma'}(\eta, \eta', n|n') $ the Wigner D-matrix related to $W(q,$ $n|n')$ as
	\begin{eqnarray}\label{eq:Wig_D_1}
		D_{\sigma\sigma'}(\eta, \eta', n|n')  &=& \langle q, \sigma'
		, n| U(W(q, n|n')) |q, \sigma, n\rangle  \nonumber \\
		&=& \langle q, \sigma', n |q, \sigma, n'\rangle
	\end{eqnarray}	 
	The second equality of Eq.(\ref{eq:Wig_D_1}), in which we made use of Eq.(\ref{eq:lg_tran_1}),  can be particularly useful in calculations.
	
	Finally, let's dicuss the choice of  momentum $q$  in Eq.(\ref{eq:gener_tran_final}) and Eq.(\ref{eq:Wig_D_1}).   $q$ is similar to the standard momentum in instant form. But unlike instant form, $q$ is not  fixed to be any specific value, also independent of $p$ and $\Lambda$. However, we can make certain choice of $q$ to make calculations more easily.  For example, we can choose $q=p$ to ensure that a general  Lorentz transformation and its corresponding little group transformation  act on the same momentum. Reflecting on the rapidities, this is also equivalent to $\eta=\ln\frac{n\cdot p}{m}$ and $\eta'=\ln\frac{n'\cdot p}{m}$.

	


	\subsection{Computation of Wigner D-matrix/d-matrix} 
	\label{subsec:wig_d}
	
	In the last subsection, we have reduced the general Lorentz transformation on the state $|p,\sigma, n\rangle$ to little group transformation on $|q,\sigma, n\rangle$ with $q$ being any momentum.  However, we  adopt the choice of $q=p$ following the example  at the end of the last subsection.  
	The remaining task then is to compute the resulting Wigner D-matrix.  
	In Eq.(\ref{eq:Wig_D_1}), we derived the formula for Wigner D-matrix,  which involves two different reference vectors($n$ and $n'$).   Since one of the two null vectors (call it $n'$) cannot be along $-z$ direction that canonically defines the front form,  Lorentz transformations defined by $n'$ cannot be computed straightforwardly using Eq.(\ref{eq:stl_1}). Here we will give a method on how to calculate them.   
	

	The basic strategy is to set up two coordinate systems $S$ and $S'$, so that $n$ in $S$ is $n^\mu=(1, 0, 0, -1)=n^\mu_S$, while $n'$ in $S'$ is also $n'^\mu_{S'}=(1, 0, 0, -1)=n^\mu$. We fix $S$, but allow $S'$ to rotate, with $n'$ and related states along with it.  In this way  we can compute $L(p,k,n')$ in $S'$ as $L^{S'}(p,k,n'_{S'})$ using the standard method of front form.   
	
	To achieve the final  goal of computing $L(p,k,n')=L^S(p,k,n')$ in $S$, for which we will drop the superscript $S$,  we notice that $n'$ is defined by its angular relation to $n$, with angle $(\theta, \varphi)$ relative to $n$.   Based on this observation, we first set  $S'$  to be  aligned with $S$ perfectly and compute $L^{S'}(p, k, n'_{S'})$,   then  we rotate System $S'$ around the origin with angle $(\theta, \varphi)$.   Then  $n'$ will obtain its intended value. $L^{S'}(p, k, n'_{S'})$ followed by the rotation of  $(\theta, \varphi)$ gives precisely $L(p,k,n')$. 
	The full $L(p, k, n')$(in $S$) is thus computed as
	\be\label{eq:stl_nprime_1}
	L(p,k,n') = R(\varphi, \theta,\varphi')L^{S'}(p,k,n'_{S'})
	\ee 
	$ R(\varphi, \theta,\varphi')$ is a chained rotation for rigid body   in z-y-z convention, that rotates $S'$ to have angle $(\theta, \varphi)$ relative to $S$: 
	\be\label{eq:rigid_rotation}
	R(\varphi, \theta,\varphi') = e^{iJ_3\varphi}e^{iJ_2\theta}e^{iJ_3\varphi'}. 
	\ee
	
	$L^{S'}(p, k,n'_{S'})$ is the standard Lorentz transformation within $S'$. 
	To compute $L^{S'}(p,k,n'_{S'})$ we need to know $p$ by components in $S'$. This can be done in analogue to $p$ in $S$, which is  $p^\mu=(\underset{+}{p\cdot n}, \underset{-}{p\cdot \bar n}, \underset{1}{p^1}, \underset{2}{p^2})$, with  $\bar n = (1, 0, 0, 1)$.  Replacing $n$ with $n'$ and $\varphi$ with $\varphi'$, we get 
	\begin{eqnarray}\label{eq:pS_prime}
		(p^{S'})^\mu=(\underset{+}{p\cdot n'}, \underset{-}{p \cdot \bar n'}, \underset{1}{(p^{S'})^1}, \underset{2}{ (p^{S'})^2}) 
	\end{eqnarray}
	$(p^{S'})^1$ and $(p^{S'})^2$ are determined by $p^1$, $p^2$ and $\theta$, $\varphi$, $\varphi'$, but it's generally complex. So we won't write them down here.  Furthermore, $n'^{S'}=(1,0,0,-1)=n$; $k^{S'}=k=(m, 0,0,0)$, because it's at rest. Putting all the elements together,   $L^{S'}(p,k,n'_{S'})$ can be  calculated as 
	\be\label{eq:stl_S}
	L^{S'}(p,k,n'_{S'}) = L(p^{S'}, k, n)
	\ee 
	Plugging the above equation into Eq.(\ref{eq:stl_nprime_1}), we then have 
	\be\label{eq:stl_nprime}
	L(p,k,n') = R_{n|n'}(\varphi, \theta,\varphi')L(p^{S'},k,n)
	\ee 
	The above formula  enables us to use the standard tools of front form to calculate the Lorentz transformation.   Notice we also added the $n|n'$ label under $R(\varphi, \theta, \varphi')$ to signify the role of $R$ in  changing the reference vector. 
	
	Next we apply those results to  one particle state and have
	\be\label{eq:stb_3}
	|p^{S'},\sigma,n\rangle &=& U(L(p^{S'},k,n))|k,\sigma\rangle
	\ee
	and
	\be\label{eq:stl_nprim_state}
	|p,\sigma, n'\rangle &=&  U(L(p,k,n'))|k,\sigma\rangle  \nonumber  \\
	&=&U(R_{n|n'}(\varphi, \theta,\varphi')L(p^{S'},k,n))|k,\sigma\rangle  \nonumber \\
	\Rightarrow |p,\sigma, n'\rangle   &=& U(R_{n|n'}(\varphi, \theta,\varphi')) |p^{S'},\sigma,n\rangle 
	\ee
	
	We're finally able to obtain the formula for  the Wigner's D-matrix that's calculable in front form.  Plugging Eq.(\ref{eq:stl_nprim_state}) into Eq.(\ref{eq:Wig_D_1}) and taking $q=p$, 
	the Wigner's D-matrix becomes
	\be\label{eq:Wigner_D_2}
	D_{\sigma \sigma'}(\eta, \eta',n|n')&=& \langle p,\sigma', n|p,\sigma,n' \rangle   \\
	& =& \langle p,\sigma', n|U(R_{n|n'}(\varphi,\theta,\varphi'))|p^{S'},\sigma,n\rangle. \nonumber 
	\ee

	We can further factor out  the dependence on the azimuthal angles $\varphi$ and $\varphi'$ in Eq.(\ref{eq:Wigner_D_2}) by defining $p_{0}$ and $ p^{S'}_{0}$ (the same way follows for $\Lambda p$)  as 
	\begin{equation}
		p_{0} =  e^{-iJ_3\varphi}p \ \ \ \ \    p^{S'}_{0} = e^{iJ_3\varphi'}  p^{S'}
	\end{equation}
	Applying them to the states in Eq.(\ref{eq:stl_nprim_state}), we get 
	\begin{eqnarray}
		e^{iJ_3\varphi'}|p^{S'}, \sigma, n\rangle &=&  e^{+i\sigma \varphi'} |p^{S'}_0, \sigma, n\rangle   \\ 
		\langle p, \sigma', n| e^{iJ_3\varphi} &=&  \langle  p_0, \sigma', n| e^{+i\sigma' \varphi} \nonumber 
	\end{eqnarray}
	Plugging into Eq.(\ref{eq:Wigner_D_2}), we then get
	\be\label{eq:Wigner_D_3}
	D_{\sigma \sigma'}(\eta, \eta',n|n') = e^{i\sigma'\varphi}e^{i\sigma\varphi'}d_{\sigma\sigma'}(\eta, \eta',n|n')
	\ee
	with 
	\be\label{eq:wig_d}
	d_{\sigma\sigma'}(\eta, \eta',n|n')= \langle p_0,\sigma', n|e^{iJ_2\theta}| p^{S'}_0,\sigma,n \rangle.
	\ee 
	$\theta$ is the angle between $n$ and $n'$. 

	The inverse of  $d_{\sigma' \sigma}(\eta, \eta',n|n')$ is:
	\be
	d^{-1}_{\sigma \sigma'}(\eta, \eta',n|n')&=&d_{\sigma' \sigma}(\eta', \eta,n'|n) \nonumber \\
	& =& \langle p^{S'}_0,\sigma, n|e^{-iJ_2\theta}|p_0,\sigma',n\rangle \nonumber 
	\ee
	
	We can also write the angle dependence explicitly in $d_{\sigma\sigma'}$ and $D_{\sigma\sigma'}$ as
	\be
	D_{\sigma\sigma'}(\eta, \eta',n|n') &=& D^{n|n'}_{\sigma \sigma'}(\eta, \eta', \theta, \varphi, \varphi') 
	\ee
	In this parameterization,  we can see clearly how the angle dependence changes under inversion: 
	\be\label{eq:wig_d_inv}
	D^{n'|n}_{\sigma'\sigma}(\eta, \eta', \theta,  \varphi, \varphi')&=&D^{n|n'}_{\sigma \sigma'}(\eta', \eta, -\theta, -\varphi, -\varphi') 
	\ee

	Finally,  the Wigner D-matrix goes to the zero-moment -um limit when  $\eta', \eta \rightarrow 0$.  In taking the limit, $|p,n,\sigma\rangle \rightarrow |k,\sigma\rangle$, the dependence on reference vector drops out.
	The  Wigner d-matrix becomes
	\be
	\lim_{\eta, \eta'\rightarrow 0} d_{\sigma\sigma'}^{n|n'}(\eta, \eta', \theta) \equiv d_{\sigma\sigma'}(\theta)
	\ee
	This is exactly  the conventional Wigner d-matrix that's defined at the rest frame.  The correct zero-momentum limit serves as a good cross check for our results.

	\section{Application on Spin-1 Particle}
	\label{sec:spin-1}
	
	\subsection{Wigner D-matrix for Spin-1 Particle}
	
	In this section    we will apply  the results in Sec.(\ref{sec:construction}) to  spin-1 particle and compute its Wigner D-matrix.
	
	Wigner D-matrix is essentially defined by two different null vectors, which we  choose as $n=(1,0,0,-1)$ and $n'^\mu=n_p\equiv (1, -\frac{\vec{p}}{|\vec{p}|})$ here.  $p$ is the momentum that the little group acts on, with $p^\mu = (p^0, |\vec p|\sin\theta \cos \varphi, |\vec p|\sin\theta\sin\varphi,$ $|\vec p|\cos\theta)$.  Our main task is to compute Wigner d-matrix from $n$ to $n_p$ using  Eq.(\ref{eq:wig_d}),  for which we  need to compute the polarization vectors defined by $n_p$ and $n$ respectively, as well as $e^{iJ_2\theta}$. 
	
	The polarization vectors for reference vector $n$ -- $\epsilon^\mu_\sigma(p,n)$ -- can be computed  by taking a standard Lorentz transformation (Eq.(\ref{eq:stl_1})) on polarization vectors at rest frame $\epsilon^\mu_\sigma(k)$.  
	In light-cone coordinates, $\epsilon^\mu_\sigma(k)$ for $\sigma =\pm$ and $\sigma =0$ are 
	\begin{equation}
		\epsilon_{\pm}^{\mu}(k) = \left(\begin{array}{c} 
			0 \\
			0\\
			\frac{1}{\sqrt{2}} \\
			\frac{\pm i}{\sqrt{2}}
		\end{array}\right) 
		\begin{array}{c} 
			+ \\
			-\\
			1\\
			2 \end{array}
		\  \  \ \ \ \  \ \ \ \  
		\epsilon_0^{\mu}(k) = \left(\begin{array}{c} 
			2 \\
			0\\
			0 \\
			0
		\end{array}\right) 
		\begin{array}{c} 
			+ \\
			-\\
			1\\
			2 \end{array}
	\end{equation}
	Transforming them to momentum $p$ along reference vector $n$  we get $\epsilon^\mu_\sigma(p,n)$
	\be\label{eq:pol_p_n}
	\epsilon_{\pm}^{\mu}(p,n)  &=& e^{\pm i\varphi}
	\left(\begin{array}{c} 
		0 \\
		\frac{\sqrt{2}p^T e^{\mp i\varphi}}{p^+}\\
		\frac{1}{\sqrt{2}} \\
		\frac{\pm i}{\sqrt{2}}
	\end{array}\right) 
	\begin{array}{c} 
		+ \\
		-\\
		1\\
		2 \end{array} \\
	\epsilon_0^{\mu}(p,n) &=& \frac{p^{\mu}}{m}+\left(\begin{array}{c} 
		0\\
		\frac{2m}{p^+} \\
		0 \\
		0
	\end{array}\right) 
	\begin{array}{c} 
		+ \\
		-\\
		1\\
		2 \end{array} \nonumber 
	\ee
	Setting $\varphi=0$,  we obtain $\epsilon^\mu_\sigma(p_0,r)$.  
	
	Like the massless case, polarization vectors for massive vector boson are also characterized by a null reference vector.   In fact, after taking Goldstone equivalence by scalarizing $\frac{k^ \mu}{m}$ into the corresponding Goldstone ``wave function", both the remnant longitudinal polarization vector $\tilde \epsilon_0(p,n) = \epsilon_0(p,n)-\frac{k}{m}$  and the transverse polarization vectors satisfy the conditions of $n\cdot \epsilon(p,n)=0$.   It turns out this special null direction indicates that  polarization vectors for massless and massive vector boson in front form are equivalent to those in spinor-helicity formalism.  This connection is explained in details in  Appendix(\ref{sec:append}).

	The polarization vectors for $p$ and $n_p$ are labeled as  $\epsilon^\mu_\sigma(p,n_p)$.  Following the method in Sec.(\ref{subsec:wig_d}), we need first to  calculate $\epsilon^\mu_\sigma(p^{S'}, n)$.   The corresponding polarization vector of $\epsilon^\mu_\sigma( p,n_p)$ at system $S'=S_p$.  $p^{S_p}$ by components is 
	\[ p^{S_p}=  (\underset{+}{p\cdot n_p}, \underset{-}{p\cdot \bar n_p}, \underset{1}{0}, \underset{2}{0})^T=  (\underset{+}{p^0+|\vec p|}, \underset{-}{p^0-|\vec p|}, \underset{1}{0}, \underset{2}{0})^T.
	\] 
	Similar to $\epsilon^\mu_\sigma(p,n)$, $\epsilon^\mu_\sigma(p^{S_p}, n)$ can be computed either by Lorentz transformation from $\epsilon(k)$, or by imposing the conditions of $p^{S_p} \cdot \epsilon(p^{S_p}, n)=0$ and $n\cdot \epsilon(p^{S_p}, n)=0$ for $\epsilon_\pm$ and $\tilde \epsilon_0=\epsilon_0-\frac{p^{S_p}}{m}$.  The result is 
	\be\label{eq:pol_vec_til_p_n}
	\epsilon_{\pm}^{\mu}( p^{S_p} ,n) &=& e^{\pm i\varphi_p}\left(\begin{array}{c} 
		0 \\
		0\\
		\frac{1}{\sqrt{2}} \\
		\frac{\pm i}{\sqrt{2}}
	\end{array}\right)  
	\begin{array}{c} 
		+ \\
		-\\
		1\\
		2 \end{array}  
	= \epsilon^\mu_\pm(p, n_p)^{S_p}		 \\
	\epsilon_0^{\mu}(p^{S_p},n) &=& \frac{(p^{S_p})^{\mu}}{m}+ \left(\begin{array}{c} 
		0\\
		\frac{2m}{p^0+|\vec{p}|} \\
		0 \\
		0
	\end{array}\right)
	\begin{array}{c} 
		+ \\
		-\\
		1\\
		2 \end{array} = \epsilon^\mu_0(p, n_p)^{S_p}  \nonumber 
	\ee
	Again, we set $\varphi_p=0$ and get $\epsilon_\sigma^\mu(p^{S_p}_0, n)$.  
	We can also obtain $e^{iJ_2\theta}$ in light-cone coordinates as
	\be\label{eq:e_J2_lc}
	&& \  \  \  \  \ \   \begin{array}{cccc} \ \ \ \ +&\  \  \  \  \  \  -&\  \ \    \ \ \ 1& \  \  \  \  \ 2 \end{array}  \n \\
	e^{iJ_2\theta}&=& \begin{array}{c}
		+ \\
		-\\
		1\\
		2\end{array}
	\left(
	\begin{array}{cccc}
		\frac{1+\cos\theta }{2}& \frac{1-\cos\theta}{2} & -\sin\theta & 0 \\
		\frac{1-\cos\theta}{2} & \frac{1+\cos\theta}{2} & +\sin\theta & 0 \\
		\frac{\sin\theta}{2} & -\frac{\sin\theta}{2} &    \cos\theta & 0 \\
		0 & 0 & 0 &1 
	\end{array}
	\right) 
	\ee
	
	Now we can plug Eq.(\ref{eq:pol_p_n}), Eq.(\ref{eq:pol_vec_til_p_n}) and Eq.(\ref{eq:e_J2_lc}) into  Eq.(\ref{eq:wig_d}),  the Wigner d-matrix elements are then   calculated as 
	\be\label{eq:wig_d_spin-1}
	d_{\sigma\sigma'}(n|n_p)  &=& \langle p_0, \sigma', n|p_0,\sigma, n_p\rangle \n \\
	&=&  \langle p_0, \sigma', n| e^{iJ_2\theta}|p_0^{S_p},\sigma, n\rangle \n \\
	\Rightarrow	d_{\sigma\sigma'}(n|n_p)&=& \frac{1}{N^{\sigma\sigma'}} \epsilon^{*}_{\sigma'\mu}(p_0,n)(e^{iJ_2\theta})^\mu_{\ \nu}\epsilon^\nu_\sigma(p_0^{S_p},n) 
	\ee
	$N^{\sigma\sigma'}$ is the normalization constant. $N^{\sigma\sigma'}=-1$ if $\sigma = \sigma'$; $N^{\sigma\sigma'}=1$ if $\sigma \neq \sigma'$.   
	Plugging  the polarization vectors into the above equation and express the above equations in terms of $\eta=\ln \frac{n\cdot p}{m}, \eta_p=\ln\frac{n_p\cdot p}{m}$ and $\theta$ only, we have
	\be\label{eq:wig_d_vec}
	d_{1,1}^{n|n_p}(\eta, \eta_p, \theta) &=& d_{-1,-1}^{n|n_p}(\eta, \eta_p,\theta) = 1-e^{-(\eta+\eta_p)}\frac{1-\cos\theta}{2} \n \\
	d_{-1,1}^{n|n_p}(\eta, \eta_p,\theta) &=&d_{1,-1}^{n|n_p}(\eta, \eta_p,\theta)=  e^{-(\eta+\eta_p)}\frac{1-\cos\theta}{2}\n\\
	d_{1,0}^{n|n_p}(\eta, \eta_p,\theta)&=&d_{-1,0}^{n|n_p}(\eta, \eta_p,\theta)=-\frac{1}{\sqrt{2}}e^{-\eta}\sin\theta  \\
	d_{0,1}^{n|n_p}(\eta, \eta_p,\theta) &=&d_{0,-1}^{n|n_p}(\eta, \eta_p,\theta)=\frac{1}{\sqrt{2}}e^{-\eta}\sin\theta \n  \\
	d_{0,0}^{n_p|n}(\eta, \eta_p,\theta) &=& 1-e^{-(\eta+\eta_p)}(1-\cos \theta) \n
	\ee
	
	This is our final formula of Wigner d-matrix for massive spin-1 particle in front form, which gives how polarization vectors transform  under the little group from reference vector $n$ to $n_p$.   By inversion (Eq.(\ref{eq:wig_d_inv})), it also gives the Wigner d-matrix from  $n_p$ to $n$.   Noticing that $n$ is solely defined as the angle relative to $n_p$,  it's then clear that we have derived the  Wigner d-matrix from  $n_p$ to any direction represented by $n$.   Based on Eq.(\ref{eq:wig_d_vec}), we can also derive the general Wigner d-matrix between any two reference vectors. Dubbing them $n$  and $n''$,  the formulae for Wigner d-matrix and the little group transformation are respectively:
	\be
	d^{n''|n}_{\sigma \sigma''}&=&\sum_{\sigma'}d^{n''|n_p}_{\sigma\sigma'}d^{n_p|n}_{\sigma'\sigma''} \\
	L(p,n''|n)&=&L(p,n''|n_p)L(p, n_p|n) 
	\ee
	
	Now let's analyze  Eq.(\ref{eq:wig_d_vec})'s limits at $\eta,\eta_p\rightarrow 0$ and $\eta, \eta_p\rightarrow \infty $, corresponding to the zero momentum limit and the massless limit respectively.  
	
	First, taking $\eta, \eta_p \rightarrow 0$,  Eq.(\ref{eq:wig_d_vec})  reduces to the Wigner d-matrix   at rest frame, thus giving the correct zero momentum limit.  
	
	Second,  we study the massless limit by taking $\eta,\eta_p\rightarrow \infty$. In this limit, the diagonal elements of the Wigner d-matrix become  1, while all non-diagonal elements go to $0$, i.e. $d_{\sigma,\sigma'}^{n|n_p}\rightarrow \delta_{\sigma\sigma'}$.  However, we need to be careful when approaching this limit, keeping in mind that what we're really computing is  how polarization vectors transform under little group: 
	\begin{equation}\label{eq:lit_tran_spin_1}
		\epsilon^\mu_{\sigma'}(p,n) \rightarrow \epsilon^\mu_\sigma(p,n_p) =  \sum_{\sigma'}d^{n|n_p}_{\sigma\sigma'}\epsilon^\mu_{\sigma'}(p,n).
	\end{equation}
	The proper massless limit is to take  $m\rightarrow 0$ for the above equation.  This subtlety is crucial, since the longitudinal polarization $\epsilon^\mu_0\rightarrow \infty$ as $m\rightarrow 0$, which cancels $d_{\pm, 0}$. Thus in the massless limit, the nontrivial components are from  $\sigma = \pm 1$ to $\sigma'=0$, giving
	\begin{equation}
		\lim_{\eta,\eta\rightarrow \infty }d_{\pm,0}^{n|n_p}\cdot \epsilon_0=-\frac{\sin\theta}{\sqrt{2}} \frac{p^\mu}{n\cdot p}
	\end{equation}

	Plugging it into Eq.(\ref{eq:lit_tran_spin_1}), we obtain that a transverse polarization vector defined at $n_p$ transforms under little group  as 
	\be
	\epsilon^\mu_\pm(p,n) &\longrightarrow & \epsilon^\mu_\pm(p,n_p)=\epsilon^\mu_{\pm}(p,n)  - \frac{\sin\theta}{\sqrt{2}} \frac{p^\mu}{n\cdot p},
	\ee
	giving  the well-known momentum shift for massless spin-1 particle. 
	
	We can also take into account of the azimuthal angle (Eq.(\ref{eq:Wigner_D_3})) by computing the D-matrix elements and get
	\be\label{eq:spin-1_mom_shift}
	\epsilon^\mu_\pm(p,n) &\rightarrow & \epsilon^\mu_\pm(p,n_p)=\epsilon^\mu_{\pm}(p,n)  - \frac{\sin\theta e^{\pm i \varphi_p}}{\sqrt{2}} \frac{p^\mu}{n\cdot p} 
	\ee
	
	In contrast to the traditional result, our method not only gives the momentum shift, but also the precise coefficients of the shift in terms of the angular parameters $\theta$ and $\varphi$.  Even more importantly, our result gives a physical interpretation for the shift:  defining   a massless particle  as the infinite boost limit of a massive particle through a null vector $n$, the momentum shift encodes the difference of approaching the limit through two different reference vectors, parameterized by the angles between them. 
	

	\subsection{Match with Wigner d-matrix from Little Group Algebra}

	So far we have successfully constructed  unitary representation of Lorentz group with manifest massive-massless continuation.  However,  in our construction we define little group transformation directly (Eq.(\ref{eq:lg_def_1})), without resorting the Lie algebra.  So  it's not clear how our construction (e.g. Eq.(\ref{eq:Wigner_D_3}) and Eq.(\ref{eq:wig_d})) is related to group contraction  in Eq.(\ref{eq:grp_contr}).  Moreover, there seems to be a mismatch of parameters between the Wigner D-matrices in our construction and  the conventional one.  A transformation under $SO(3)$ is parameterized by 3 Euler angles,  meanwhile Wigner D-matrix in Eq.(\ref{eq:Wigner_D_3}) has 5 parameters. So it's not even clear if the two methods are equivalent.  In this section we are resolving this problem by  proving the two approaches are indeed equivalent and finding the relations between the parameters.

	First, we implement group contraction from Pauli-Luban-ski pseudovector $S_\mu=\epsilon_{\mu\nu\rho\sigma}M^{\nu\rho}P^\sigma$, which commutes with $P^\mu$.   For simplicity,  we set the momentum  $p^\mu=(p^0, 0,0, |\vec{p}|$) as the eigenvalue of $P^\mu$.  The components of  $S_{\mu}$ are 
	\be
	S_0&=&|\vec{p}|J^3 \ \ \ S_3 = p^0J^3 \\
	S_1 &=& p^0J^1 + |\vec{p}| K^2 \ \ \   S_2 = p^0J^2 - |\vec{p}| K^1 \nonumber
	\ee
	$J_i$ and $K_i$ are generators of rotations and boosts respectively.  We can then choose $\{J_3\equiv \frac{1}{m}\sqrt{-S_0^2+S_3^2}, S_1,S_2\}$ as the basis. Their commutation relations are
	\begin{equation}\label{eq:grp_contr_S}
		[S_1,S_2]=im^2J_3\  \   \ [S_2,J_3]=iS_1\ \ \  [J_3,S_1]=iS_2 
	\end{equation}
	which are precisely the group contraction of $SO(3)$ in Eq.(\ref{eq:grp_contr}) if we identify $m=\epsilon$. 
	
	Moreover, $S_1/m$ and $S_2/m$ play the role of $J_1$ and $J_2$ respectively in the $SO(3)$ Lie algebra.  In analogue to the case of rest frame,  the Wigner D-matrix with a general momentum  can then be defined as 
	\be
	D^{n_p}_{\sigma\sigma'}= \langle p, \sigma', n_p| W_3(\varphi)W_2(\alpha)W_3(\varphi') |p,\sigma, n_p \rangle
	\ee
	with $W_2(\alpha)\equiv e^{i\frac{S_2}{m}\alpha}$ and $W_3(\varphi)\equiv e^{iJ_3\varphi}$.
	Factoring out the azimuthal angle dependence, we get the Wigner d-matrix 
	\be\label{eq:wig_d_lg}
	d_{\sigma\sigma'} = \frac{1}{N^{\sigma\sigma'}} \cdot ( \epsilon_{\sigma'}(p_0,n_p) \cdot   W_2(\alpha_2) \cdot  \epsilon_\sigma(p_0,n_p) )
	\ee
	As in Eq.(\ref{eq:wig_d_spin-1}), $N^{\sigma\sigma'}=-1$ when $\sigma = \sigma'$;  $N^{\sigma\sigma'}=1$ when $\sigma\neq \sigma'$.  
	Thus in order to compute the Wigner D-matrix, we need know the form of $S_2$.  Before the calculations, however, notice  $W_2(\alpha)\rightarrow \infty$ as $m\rightarrow 0$, making it seems there is no smooth massless limit.  This blow-up can be avoided if  $\alpha_{2}\propto m$ as $m\rightarrow 0$, which will be checked with the final results.   
	
	In light-cone coordinates, $S_2$ is calculated to be  
	\be
	&& \  \  \  \  \ \  \  \  \   \  \begin{array}{cccc} +&\  \  \  \  \  \  \ \  \ -&\  \ \  \ \  \ \ \ \ \   \ \ 1& \  \  \ \   \  \  \  \ 2 \end{array}  \n \\
	(S_2)^{\mu}_{\nu}&=& \begin{array}{c}
		+ \\
		-\\
		1\\
		2\end{array} \left[ \begin{array}{cccc}
		0 & 0& i(p^0+|\vec{p}|) & 0\\
		0 & 0& -i(p^0-|\vec{p}|) &  0 \\
		\frac{-i(p^0-|\vec{p}|) }{2}& \frac{+ i(p^0+|\vec{p}|)}{2} & 0& 0\\
		0&0&0&0\\
	\end{array}\right] \n
	\ee
	Furthermore, we find $(S_2)^3=m^2S_2$, $(S_2)^4=m^2S_2^2$, which means that for general $n=1,2,...$,  we have 
	\be
	\left(\frac{S_2}{m}\right)^{2n+2}= \left(\frac{S_2}{m}\right)^2  \  \ \ \
	\left(\frac{S_2}{m}\right)^{2n+1}= \frac{S_2}{m}
	\ee
	$e^{+i\frac{S_2}{m}\alpha_2}$ can then be written as 
	\be
	e^{+i\frac{S_2}{m}\alpha_2}=(\cos\alpha_2 -1)\left(\frac{S_2}{m}\right)^2 +i\sin\alpha_2\frac{S_2}{m}+1
	\ee
	Plugging back to Eq.(\ref{eq:wig_d_lg}), we can then obtain the Wigner d-matrix elements as
	\be\label{eq:wig_d_5}
	d_{1,1}(\alpha)&=& d_{-1,-1}(\alpha)= \frac{1+\cos\alpha}{2}\n \\
	d_{1,-1}(\alpha)&=& d_{-1,1}(\alpha) = \frac{1-\cos\alpha}{2} \n \\
	d_{1,0}(\alpha)&=& d_{-1,0}(\alpha)=+\frac{1}{\sqrt{2}}\sin\alpha  \\
	d_{0,1}(\alpha)&=& d_{0,-1}(\alpha)=-\frac{1}{\sqrt{2}}\sin\alpha \n \\
	d_{0,0}(\alpha) &=& \cos\alpha \n
	\ee
	Comparing with Eq.(\ref{eq:lg_tran_1}) and Eq.(\ref{eq:Wig_D_1}), Eq.(\ref{eq:wig_d_5}) should be equivalent to the inverse of  Eq.(\ref{eq:wig_d_vec}), i.e. $d^{n_p|n}_{\sigma'\sigma}$. 
	Therefore we should be able to  find the relations between their parameters by matching the matrix elements. 
	
	Eq.(\ref{eq:wig_d_5}) has only one parameter, while Eq.(\ref{eq:wig_d_vec})  has three. This makes the match non-trivial. 
	Now let's first match the two   $d_{0,1}$s  to be equal, which gives
	\be\label{eq:wig_alpha}
	\sin\alpha = e^{-\eta} \sin\theta= \frac{m}{n\cdot p}\sin\theta.
	\ee
	This fixes  $\alpha$, therefore the whole Eq.(\ref{eq:wig_d_5}),  in terms of $\theta$ and $\eta$.  If Eq.(\ref{eq:wig_d_5}) and the inverse of Eq.(\ref{eq:wig_d_vec}) are really equivalent,  other components must also match. To check this,  we  plug  Eq.(\ref{eq:wig_alpha}) into  $d_{1,-1}$ in Eq.(\ref{eq:wig_d_5}) and after some manipulations(see in Appendix (\ref{sec:append-2})), we prove that 
	\be\label{eq:d_1-1}
	d_{1,-1}(\alpha)= \frac{1}{2}(1-\cos\alpha)= e^{-(\eta+\eta_p)}\frac{1-\cos \theta}{2}
	\ee
	The match of other components can be checked straightforwardly.   Eq.(\ref{eq:wig_alpha}) and Eq.(\ref{eq:d_1-1}) also show that the only free, physical parameter in d-matrix is $\alpha=\arcsin(\frac{m}{n\cdot p}\sin\theta)$.  $\eta_p$ is determined by $\theta$, $\eta$. Although both $\theta$ and $\eta$ are independent, neither fully specify the Lorentz  transformation(assuming $\varphi/\varphi'$ are already fixed), only the single variable $\alpha$ after the unique combination in Eq.(\ref{eq:wig_alpha}) does.   
	
	Finally, we point out that as $m\rightarrow 0$,  $ \sin\alpha \rightarrow \alpha =  \frac{m}{n\cdot p}\theta$, so  we have $\lim_{m\rightarrow 0}\alpha \propto m$. As a result,   $\alpha$ cancels $m$ in the denominator in $W_2(\alpha)=e^{i\frac{S_2}{m}\alpha_2}$, ensuring a smooth massless limit for $W_2(\alpha)$, as promised.
	


	\section{Conclusion and Discussion}
	\label{sec:con}

	In this paper, we successfully constructed one particle states as unitary representations of Poincare group in front form with massive-massless continuation.  The state is defined by a null reference vector, also a mark of front form.  We then found  the little group transformation and subsequently  the Wigner D-matrix, can be defined as  change of reference vectors.  This allows us to derive concrete formulae (Eq.(\ref{eq:Wig_D_1}) and Eq.(\ref{eq:Wigner_D_2})) for Wigner D-matrix, which has continuous zero momentum limit and massless(infinite boost) limit.
	
	We then applied our method to massive spin-1 particle and computed Wigner D-matrix elements.  Taking  the massless limit, we not only recovered the famous momentum shift $\epsilon_s\rightarrow \epsilon_s +\xi k $, but also obtained the coefficients of the momentum shift in terms of angle parameters $(\theta, \varphi)$.   Finally we computed Wigner D-matrix directly from Lie algebra of the little group  based on the Pauli-Lubanski pseudovector. The results agree with our construction in front form by making the  identification of the two sets of parameters as in Eq.(\ref{eq:wig_alpha}).  Our approach on spin-1 particles gives us a physical interpretation of the momentum shift as the difference of  the same particle state defined through different reference vectors, thus deepening our understanding of gauge symmetry.  

	Our work also opens up many possibilities for further research.  The most immediate one is to extend our method to spin-$\frac{1}{2}$ particle, which we will explore  in our future work. 
	Considering the connection of  wave functions in front form and spinor-helicity formalism,  our formulae also have  potential of shedding light on the on-going research on massive amplitudes.   We will also study those directions in future work.  Moreover,  generalizing our method to  particles with higher spins could also be interesting. 
	
	
	As a concluding remark,  we comment on the possible objection that our approach breaks rotational symmetry by choosing a special  direction through the reference vector. This, however, is not a valid objection.  Even in the traditional instant form,  we need both $J^2$ and $J_3$ to fully specify a state with spin. They don't break rotational symmetry, because through little group and Wigner D-matrix we know how the state defined on different z directions transform to each other.  In the same way,  after Wigner D-matrix specifies how the same state defined on different null vectors transform to each other in front form,   there is no symmetry ``broken" in Lorentz group. The only difference is that, the ``rotations'' -- changes of direction - in front form correspond to linear combinations of $J_i$ and $K_i$.

	\section*{Acknowledgements}
	
	We would like to thank Stanley Brodsky for stimulating discussions a few years ago, which serves as a crucial motivation for this paper.  This work is supported by National Natural Science Foundation of China under Grant No. 12205118.
	
	\appendix

	\section{Equivalence of Polarization Vectors in Front Form and Spinor-Helicity Formalism}
	\label{sec:append}
	
	The first clue for the connection between wave functions in  front form and  spinor-helicity formalism can be found in momentum $p^\mu$ in light-cone coordinates: $p^\mu$, multiplying which with $\bar \sigma^{\mu \dot \alpha \alpha}=(\delta^{\dot\alpha \alpha}, -\vec \sigma^{\dot \alpha \alpha})$, we get 
	\begin{equation}\label{eq:p_sig}
		p^{\dot\alpha \alpha}= p\cdot \bar\sigma^{\dot \alpha \alpha} = 
		\left(\begin{array}{cc}
			p^+ &  p^L \\
			p^R &  p^- 
		\end{array}\right)
	\end{equation}
	with $p^L=p^1-ip^2$ and $p^R = p^1 + ip^2$.  Eq.(\ref{eq:p_sig}) is usually taken as a starting point of spinor-helicity formalism \cite{Witten:2003nn,Schwartz:2014sze}. Notice  that the terms of $p_{\alpha\dot\alpha}$ in Eq.(\ref{eq:p_sig}) are in one-to-one correspondence with $p^\mu$ in light-cone coordinates if written as $p^\mu = (\underset{+}{p^+}, \underset{-}{p^-}, \underset{L}{p^L}, \underset{R}{p^R})$.  Also since  $p^+$ can be written as $p^+=p\cdot n$  with $n^\mu=(\underset{+}{0}, \underset{-}{2},\underset{L}{0},\underset{R}{0})$,  momentum $p$ in spinor representation is characterised by the null vector $n$.  
	Since  wave functions for any spin can be constructed from the spinors that come from the decomposition of $p^{\dot \alpha\alpha}$,  wave functions in spinor-helicity formalism are characterized by a special null direction too, just as front form. Thus we establish a connection between spinor formalism to front form, which is also defined with a special null vector. 
	

	Now we turn to spin-1 particle. For massless spin-1 particle, the polarization vectors in front form  can be fixed by the two conditions of 
	\begin{eqnarray}\label{eq:eps_con}
		p\cdot \epsilon_\sigma = 0 \ \ \ \ \       n\cdot \epsilon_\sigma =0,
	\end{eqnarray}
	in addition to the normalization and orthogonal conditions.  The latter two conditions don't fix any degree of freedom(d.o.f.), therefore the number of  d.o.f. for $\epsilon^\mu$ after the conditions is $4-2=2$, corresponding to the two transverse polarizations.  
	
	This construction can be converted to spinor formalism.  Making use of the identity $g_{\mu\nu} = \frac{1}{2}\text{tr}(\bar \sigma_\mu \sigma_\nu)$ (or equivalently $g_{\mu\nu} = \frac{1}{2}\text{tr}(\sigma_\mu \bar \sigma_\nu)$) and decomposing $p$ and $n$  into outer products of spinors as
	\begin{equation}\label{eq:p_spin}
		p^{\dot \alpha \alpha} = \tilde \lambda^{\dot \alpha} \lambda^{ \alpha} 
		\  \ \ \ \  n^{\dot \alpha \alpha} = \tilde r^{\dot \alpha}  r^{\alpha},
	\end{equation}
	the two conditions become
	\begin{equation}\label{eq:eps_con_spin}
		\lambda^\alpha \epsilon_{\alpha \dot \alpha} =0
		\ \ \ \ \ r^\alpha \epsilon_{\alpha\dot \alpha}=0.
	\end{equation}
	Those are exactly the conditions for polarization vectors in spinor-helicity formalism\cite{Schwartz:2014sze}.  Thus  polarization vectors  in front form are indeed equivalent to those in spinor-helicity formalism for massless spin-1 particle. 
	
	For massive spin-1 particle,  the conditions for polarization vectors are modified.  The transverse polarization vectors still satisfy Eq.(\ref{eq:eps_con}).   For the longitudinal polarization vector, we need first to take Goldstone equivalence and scalarize the $\frac{k}{m}$ term, then the remnant term $\tilde \epsilon_0^\mu = \epsilon_0^\mu-\frac{k^\mu}{m}$ satisfies $n\cdot \tilde \epsilon_0=0$. Therefore polarization vectors for massive vector boson also have a special null direction. 
	
	Eq.(\ref{eq:eps_con}) can also be converted to spinor formalism in the massive case.   For $p^2=m^2\neq 0$, $p^\mu$ can be decomposed as 
	\begin{equation}
		p^\mu = \tilde p^\mu + \frac{m}{2n\cdot p} n^\mu,  \ \ \ \  \tilde p^2=n^2=0
	\end{equation}
	Both $\tilde p$ and $n$ can then be multiplied by $\bar \sigma$ and decomposed into spinors as in Eq.(\ref{eq:eps_con_spin}).  Still making use of $g_{\mu\nu}=\frac{1}{2}\text{tr}(\sigma_\mu\bar\sigma_\nu)$,  the conditions turn to the spinor counter parts in Eq.(\ref{eq:eps_con_spin}). Therefore polarization vectors for massive spin-1 particle in front form are also related to that in spinor formalism.  After scalarizing the $\frac{k}{m}$ term by taking Goldstone equivalence,  the two forms  become equivalent.

		\section{Derivation of Eq.(\ref{eq:d_1-1})}
				\label{sec:append-2}

	Here we  prove Eq.(\ref{eq:d_1-1}). 
	Plugging Eq.(\ref{eq:wig_alpha}) into the $d_{1,-1}$ in Eq.(\ref{eq:wig_d_5}), we have 
	\be
	d_{1,-1}(\alpha)&=&\frac{1}{2}(1-\cos\alpha)\\
	&=& \frac{1}{2}\left(1-\sqrt{1-(\frac{mp^T}{p^+|\vec{p}|})^2}\right)\n \\
	&=&  \frac{1}{2} \frac{m^2(p^T)^2}{(p^+)^2|\vec{p}|^2}\frac{p^+|\vec{p}|}{p^+|\vec{p}|+\sqrt{(p^+)^2|\vec{p}|^2-m^2(p^T)^2}}\n
	\ee
	The expression under the square root can be further simplified as
	\be
	(p^+)^2|\vec{p}|^2-m^2(p^T)^2&=& (p^+)^2|\vec{p}|^2-m^2(p^+p^--m^2)\n \\
	&=&  (p^+)^2((p^0)^2-m^2)-m^2(p^+p^-m^2) \n \\
	&=& (p^+p^0-m^2)^2
	\ee
	plugging back in, and after some algebraic manipulations, we have 
	\be
	d_{1,-1}(\alpha)&=& \frac{1}{2}(1-\cos\alpha)\n\\
	&=& \frac{1}{2}\frac{m^2(p^T)^2}{(p^+)^2|\vec{p}|^2}\frac{p^+|\vec{p}|}{p^+(|\vec{p}|+p^0)-m^2} \n \\
	&=&  \frac{1}{2}\frac{m^2(p^T)^2}{p^+|\vec{p}|(p^0+|\vec{p}|)(|\vec{p}|+p^3)}\n\\
	\Rightarrow d_{1,-1}(\alpha)&=& \frac{m^2}{(n\cdot p)( r_p\cdot p)}\frac{(1-\cos\theta)}{2}
	\ee
	which is precisely $d_{1,-1}^{n|n_p}$ in Eq.(\ref{eq:wig_d_vec}).  From the third line to the fourth, we made use of 
	\be
	\frac{(p^T)^2}{|\vec{p}|(p^0+|\vec{p}|)(|\vec{p}|+p^3)}&=& \frac{|\vec{p}|^2-(p^3)^2}{|\vec{p}|(p^0+|\vec{p}|)(p^+-\frac{m^2}{p^0+|\vec{p}|)})} \n \\
	&=& \frac{|\vec{p}|^2-(p^3)^2}{|\vec{p}|(p^0+|\vec{p}|)(p^+-(p^0-|\vec{p}|))} \n\\
	&=&\frac{1-\cos\theta}{p^0+|\vec{p}|}
	\ee
	Thus we proved Eq.(\ref{eq:d_1-1}).

	
	\bibliographystyle{apsper}	
	\bibliography{lit-lorentz-gauge}

\end{document}